\documentstyle[12pt]{article}

\input{epsf}
\input psfig.sty

\def\kms{km s$^{-1}$} 
\def\ha{H$\alpha$}

\begin{document}

\title{Accurate Rotation Curves and Distribution of
Dark Matter in Galaxies\footnote{To appear in the Proceedings
of  XIXth Moriond Astrophysics Meeting  ``Building Galaxies: from the
 Primordial Universe to the Present'', Les Arcs, March 13-20 1999.
ed. F.Hammer et al. (Editions Frontieres, Gif-sur-Yvetter)
}}

\input{epsf}
\input psfig.sty

\def\kms{km s$^{-1}$}

\author{Yoshiaki Sofue \\
Institute of Astronomy, University of Tokyo,
Mitaka, Tokyo 181-8588, Japan.}

\maketitle

\begin{abstract}
We present high-accuracy rotation curves, which show 
a steep nuclear rise and high-velocity central rotation,
followed by a broad maximum in the disk and flat part.
We use the rotation curves to directly calculate the radial 
distribution of surface mass density, and obtain radial 
variations of the mass-to-luminosity ratio (M/L).
The M/L ratio and, therefore, the dark mass fraction (DMF)
is not constant at all, but varies within the bulge,
increases already within the disk,
and rapidly from the disk toward halo. 
In some galaxies, the DMF within the bulge increases 
inward toward the center, indicating a massive dark core. 

\end{abstract}

\section{Introduction}

Rotation curve is the principal tool to derive the axisymmetric
distribution of mass in disk galaxies in the first-order approximation.
Rotation curves have been obtained by optical and HI-line 
spectroscopy (Rubin et al 1980, 1982; Bosma 1981;
Mathewson et al 1996; Persic et al 1996). 
However, the inner rotation curves have been not thoroughly 
investigated yet, not only because the concern in these studies 
has been on the massive halo, but also for the difficulty in 
observing inside central bulges.
We have shown that the CO molecular line is useful for 
deriving central kinematics for its high concentration
in the center and low extinction
(Sofue 1996, 1997, Sofue et al 1997, 1998: Papers I to IV).
Recent CCD \ha\ line spectroscopy has also made us
available with accurate rotation curves for the inner
regions (Rubin et al 1997; Sofue et al 1998). 
In this paper, we present high-accuracy rotation curves, 
and discuss their general characteristics. 
We derive surface mass distributions, and discuss the radial 
variation of mass-to-luminosity ratio and the dark mass fraction. 

\begin{figure} 
\psfig{figure=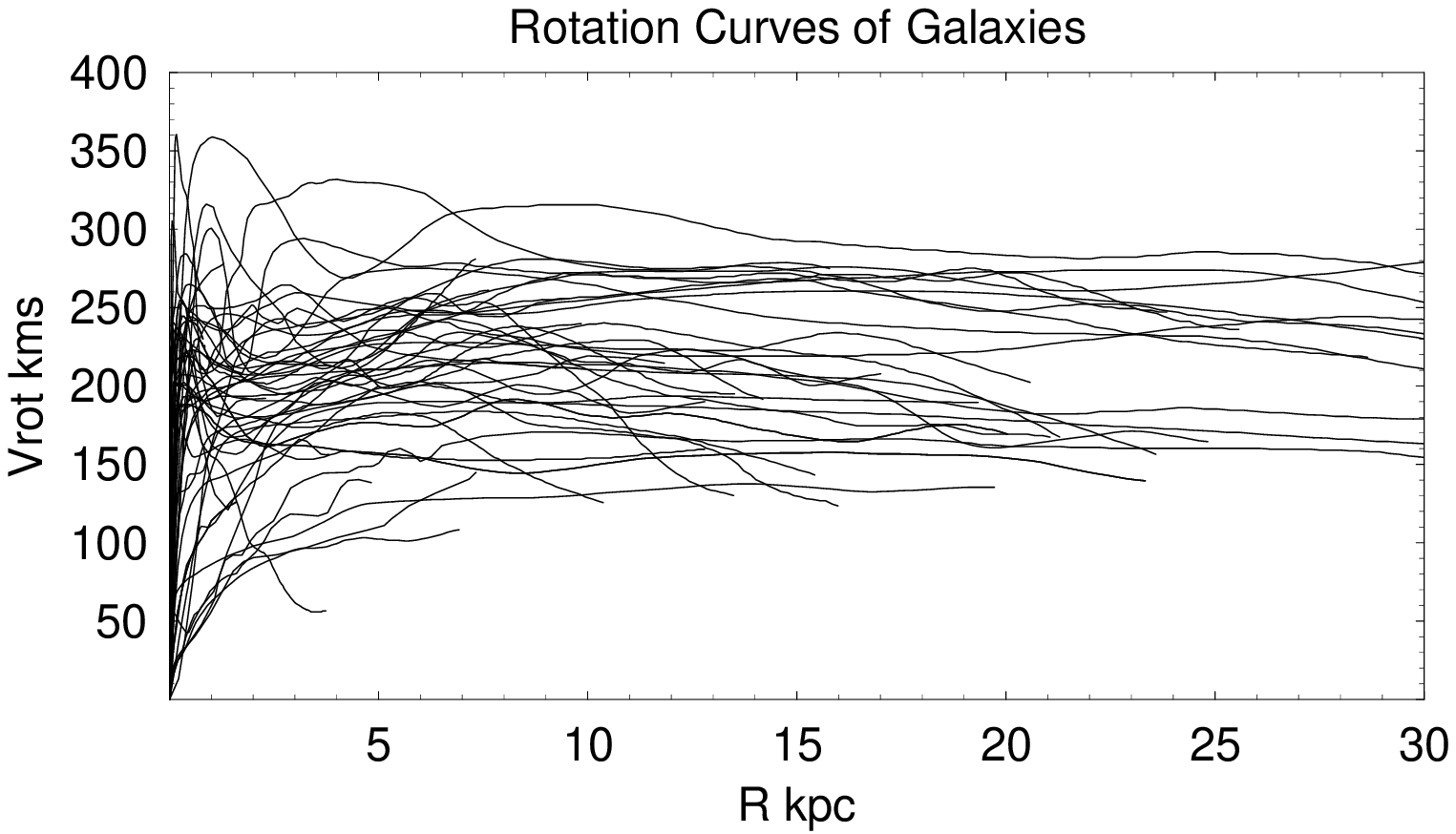,height=6cm}
Fig. 1a. Most-completely-sampled rotation curves  of Sb, Sc, SBb 
and SBc galaxies obtained by using CO, \ha\ and HI-line data. 
\end{figure}
 
\begin{figure} 
\psfig{figure=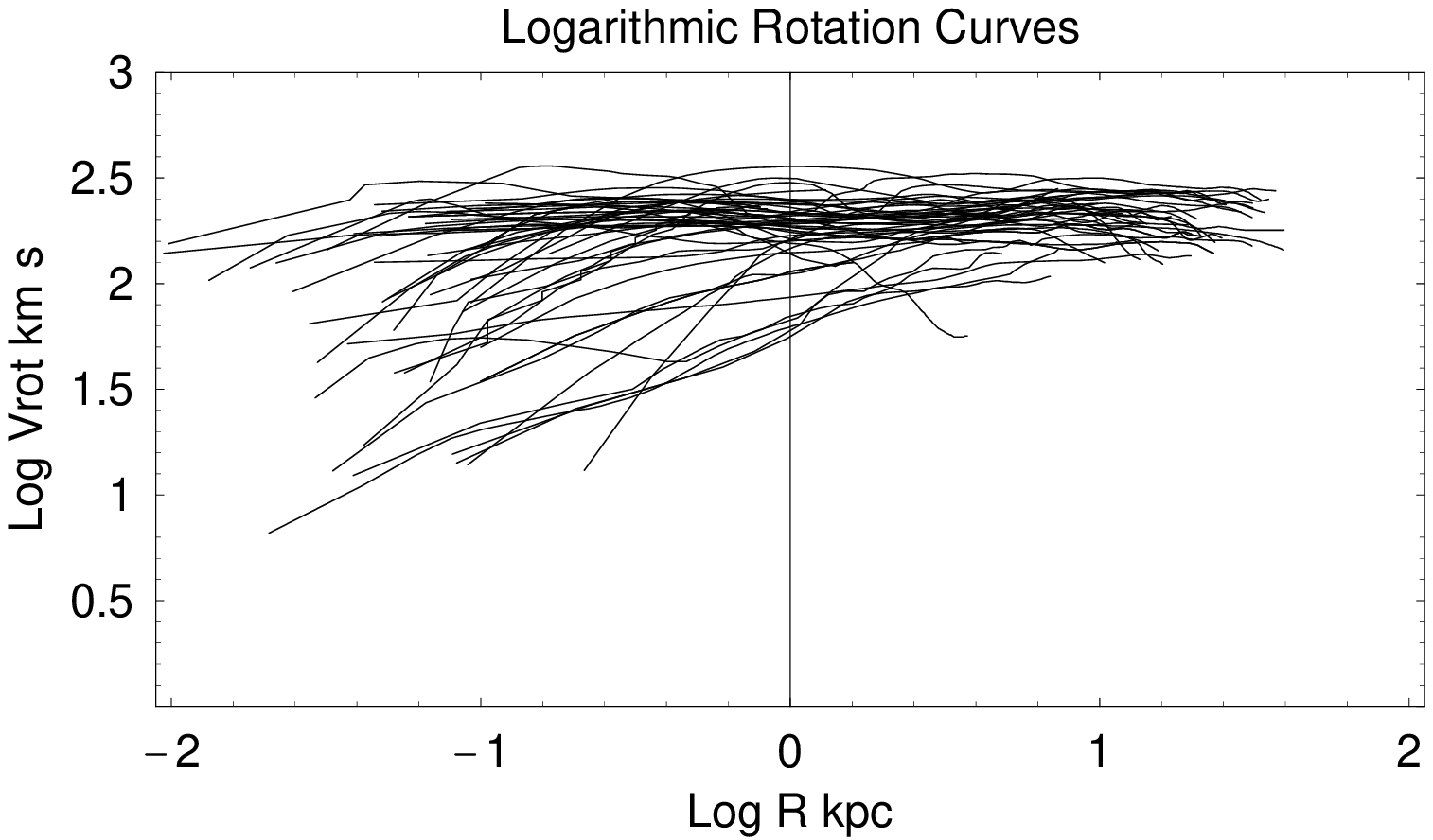,height=6cm}
Fig. 1b. The same as Fig. 1a, but in a logarithmic plot.
\end{figure}

\section{Universal Properties of Rotation Curves}

\subsection{Central-to-Outer Rotation Curves}
 
Besides the Milky Way, it has been widely believed
that inner rotation curves behave in a rigid-body fashion.
In order to clarify if such rigid rotation is common, or
galaxies have similar rotation curves to the Milky Way,
we have performed high-resolution CO-line observations.
We have also obtained CCD spectroscopy in the \ha\
and [NII] line emissions of the central regions of galaxies.
In deriving rotation curves, we applied the envelop-tracing
method from PV diagrams.
In Fig. 1a, we show the most-completely-sampled rotation
curves (Papers I - IV). 

\subsection{Logarithmic Rotation Curves}

Since the dynamical structure of a galaxy varies with the radius
rapidly toward the center, a logarithm plot
would help to overview the innermost kinematics.
In fact, logarithmic plots in Fig. 1b demonstrate the 
convenience to discuss the central kinematics. 
In such a plot, we may argue that high-mass galaxies show 
almost constant rotation velocities from the center to outer edge.
 
\subsection{Universal Properties}

We may summarize that the universal properties of  rotation curves
in Fig. 1 and 2 as follows, which are similar to those for the Milky Way.
 
\vskip 2mm

(1) Steep central rise and peak, often starting from
high velocity at the nucleus;

(2) Bulge component, often causing the central peak of rotation curve;

(3) Broad maximum by the disk; and

(4) Halo component. 
\vskip 2mm

The steep nuclear rise of rotation
is a universal property for massive Sb and Sc galaxies,
regardless the morphological peculiarities,
while less massive galaxies tend to show a rigid-body rise.
The fact that almost all massive galaxies show
the steep rise indicates that it is not due to
non-circular motion by chance.
Even if there is a bar, we have more chance to observe
shocked gas bound to the potential than high-velocity flows: 
We have more chance to observe the pattern speed than
the high-velocity flow, which would result in underestimating
the true rotation velocity.

\section{Mass-to-Luminosity Ratio and the Dark Mass Fraction (DMF)}

Once an accurate rotation curve from the center to the outer edge
is obtained, we can directly calculate the surface mass density.
One extreme case is to assume a spherical symmetry:
the rotation velocity is used to calculate the total mass involved 
within a radius, which is then used to calculate the surface mass density.
Another extreme case is to assume a thin disk:
the surface mass density can be directly calculated by using 
the Poisson equation (e.g. Binney and Tremaine 1987).
We may safely assume that the true mass distribution lies 
in between these two cases.
We have, thus, calculated the mass distribution for the galaxies
for which accurate rotation curves have been obtained.
Results for spherical and disk assumptions are found to be 
coincident usually within a factor of 1.5 to 2.
We stress that this method is not intervened by any potential models,
as widely adopted in such a method to fit calculated rotation curves
by assuming potentials (Kent 1987).

\begin{figure}
\psfig{figure=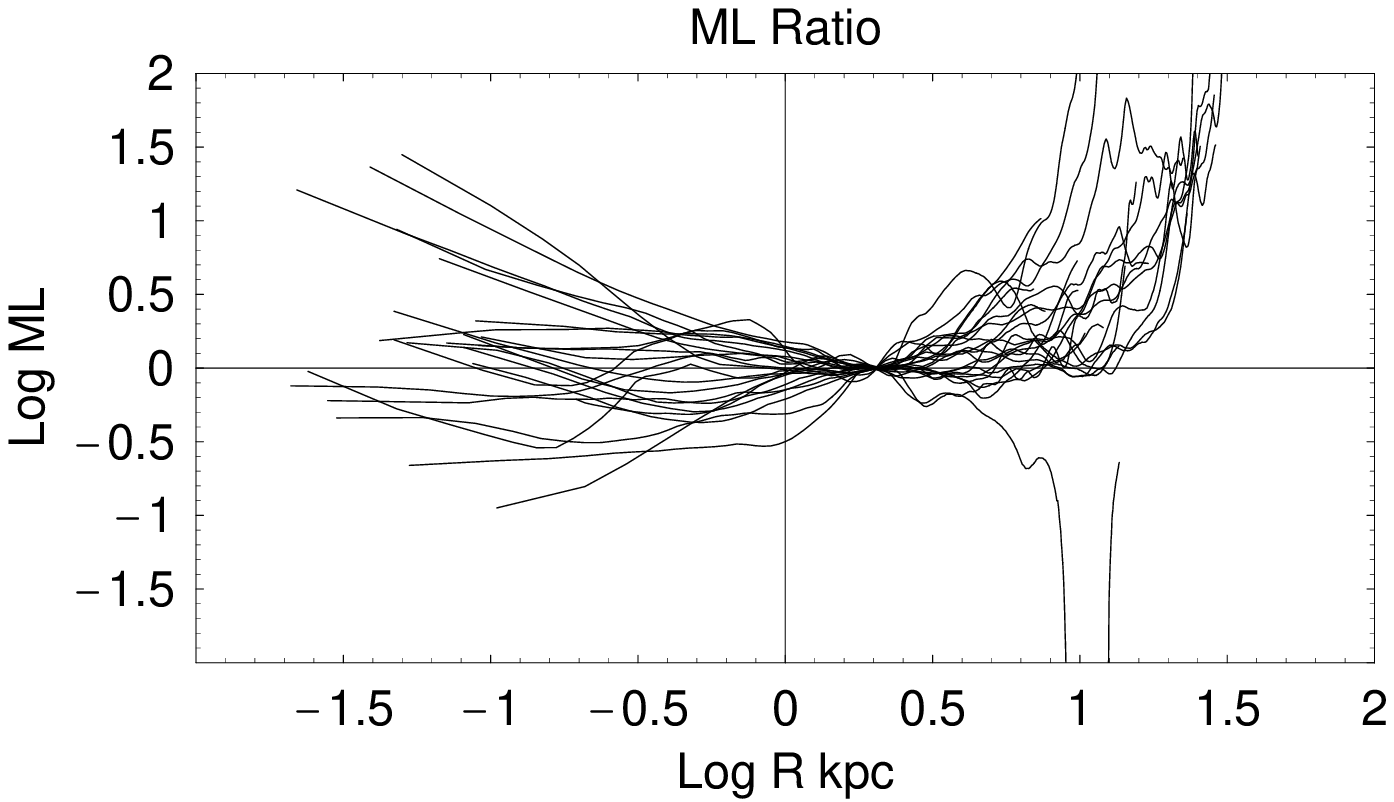,height=6cm}
Fig. 2. Radial variation of M/L ratio, as normalized to unity at 
radius 2 kpc.
M/L is not constant at all, but varies within the bulge and disk,
and steeply increases toward the halo.
\end{figure}
 
The surface mass density can be, then, directly compared with
observed surface luminosity, from which we can derive the
mass-to-luminosity ratio (M/L)
Fig. 2 shows the thus obtained radial distributions of M/L 
for a disk assumption (Takamiya and Sofue 1999).
The figure indicates that the M/L is not constant at all,
but varies significantly within a galaxy.
Since the M/L of stars will not vary so drastically, this diagram 
can be interpreted to represent the distribution of the dark mass 
fraction (DMF) for the first approximation, namely the minimal DMF.

\vskip 2mm
 
(1) M/L and DMF vary drastically within the central bulge.
In some galaxies, it increases inward toward the center,
suggesting a dark massive core.
In some galaxies, it decreases toward the center, 
likely due to luminosity excess such as due to active nuclei.

(2) M/L and DMF gradually increases from the inner disk to outer disk, 
and the gradient increases with the radius.

(3) M/L and DMF increases drastically from the outer disk toward
the outer edge, indicating the massive dark halo.
In many galaxies, the dark halo can be nearly directly
seen from this figure, where the M/L exceed ten, and sometimes
hundred.  

\vskip 8mm
\noindent{\bf References}

\def\r{\hangindent=1pc \noindent \\}
 
\r Binney, T., Tremaine, S. 1987, in Galactic Astronomy (Princeton
Univ. Press).
\r Bosma  A. 1981,  AJ  86,  1825 
\r Kent, S. M. 1987, AJ 93, 816.
\r Mathewson, D.S. and Ford, V.L., 1996 ApJS, 107, 97.
\r Persic, M.,  and Salucci, P.  1995, ApJS 99, 501.
\r Rubin  V. C., Ford  W. K., Thonnard  N. 1980, ApJ  238, 471
\r Rubin, V. C., Ford, W. K., Thonnard, N. 1982, ApJ, 261, 439
\r Rubin, V., Kenney, J.D.P., Young, J.S. 1997 AJ, 113, 1250.
\r Sofue, Y. 1996, ApJ, 458, 120 (Paper I)
\r Sofue, Y. 1997, PASJ, 49, 17 (Paper II)
\r Sofue, Y., Tomita, A.,  Honma, M., Tutui, Y. and Takeda, Y.
        1998, PASJ 50, 427. (Paper IV)
\r Sofue, Y.,  Tutui, Y., Honma, M., and Tomita, A.,
        1997, AJ, 114, 2428 (Paper III)
\r Takamiya, T., and Sofue, Y. 1999, in preparation.

\end{document}